\documentclass[a4paper,fleqn,usenatbib]{mnras}

\usepackage{mathptmx}
\usepackage{amssymb}
\usepackage[T1]{fontenc}
\usepackage{ae,aecompl}

\usepackage{graphicx}	
\usepackage{amsmath}	
\usepackage{amssymb}	
\usepackage{natbib} 

\title[Spectropolarimetric capabilities of Ca~{\sc ii} 8542~\AA \ line]{Spectropolarimetric capabilities of Ca~{\sc ii} 8542~\AA \ line}

\author[C. Quintero Noda et al.]{C. Quintero Noda,$^{1}$\thanks{E-mail: carlos@solar.isas.jaxa.jp}
T. Shimizu,$^{1}$
J. de la Cruz Rodr\'{i}guez,$^{2}$
Y. Katsukawa,$^{3}$
\newauthor
K. Ichimoto,$^{4}$
T. Anan,$^{4}$
Y. Suematsu$^{3}$
\\
$^{1}$Institute of Space and Astronautical Science, Japan Aerospace Exploration Agency, Sagamihara, Kanagawa 252-5210, Japan\\
$^{2}$Institute for Solar Physics, Dept. of Astronomy, Stockholm University, Albanova University Center, SE-10691 Stockholm, Sweden\\
$^{3}$National Astronomical Observatory of Japan, 2-21-1 Osawa, Mitaka, Tokyo 181-8588, Japan\\
$^{4}$Kwasan and Hida Observatories, Kyoto University, Kurabashira Kamitakara-cho, Takayama-city, 506-1314 Gifu, Japan\\
}

\date{Accepted XXX. Received YYY; in original form ZZZ}

\pubyear{2015}

\begin{document}
\label{firstpage}
\pagerange{\pageref{firstpage}--\pageref{lastpage}}
\maketitle

\begin{abstract}
The next generation of space and ground-based solar missions aim to study the magnetic properties of the solar chromosphere using the infrared Ca~{\sc ii} lines and the He~{\sc i} 10830 \AA \ line. The former seem to be the best candidates to study the stratification of magnetic fields in the solar chromosphere and their relation to the other thermodynamical properties underlying the chromospheric plasma. The purpose of this work is to provide a detailed analysis of the diagnostic capabilities of the Ca~{\sc ii} 8542 \AA \ line, anticipating forthcoming observational facilities. We study the sensitivity of the Ca~{\sc ii} 8542 \AA \ line to perturbations applied to the physical parameters of reference semi-empirical 1D model atmospheres using response functions and we make use of 3D MHD simulations to examine the expected polarization signals for moderate magnetic field strengths. Our results indicate that the Ca~{\sc ii} 8542 \AA \ line is mostly sensitive to the layers enclosed between $\log$ $\tau=[0,-5.5]$, under the physical conditions that are present in our model atmospheres. In addition, the simulated magnetic flux tube generates strong longitudinal signals in its centre and moderate transversal signals, due to the vertical expansion of magnetic field lines, in its edge. Thus, observing the Ca~{\sc ii} 8542 \AA \ line we will be able to infer the 3D geometry of moderate magnetic field regions.

\end{abstract}




\section{Introduction}

The solar chromosphere embodies the transition between the photosphere and the corona, two regions characterized by very different physical conditions. On one hand, the photosphere is a cool layer (around 6000 K) governed by convection, with large plasma-$\beta$ (ratio between kinetic pressure and magnetic pressure) values. On the other hand, the corona is a very hot region (more than $10^{6}$ K) where the plasma density is sufficiently small to produce very low plasma-$\beta$ values \citep[for an introduction, see][]{Stix1989}. In the chromosphere, the plasma-$\beta$ falls below unity, signalling a shift from hydrodynamic to magnetic forces as the dominant agent in the structuring of the atmosphere. Thus, we expect combined effects of magnetic field guidance and small-scale gas thermodynamics in the chromosphere, which probably leads to the complex chromospheric structures as jets, spicules, fibrils or mottles \citep[see, for instance,][]{Judge2006,Rutten2007}.

This highly dynamic and complex layer of the sun has become a major target of the most important future missions as DKIST  \citep[previously known as ATST,][]{Keil2003,Keil2011,Elmore2014}, EST \citep{Collados2013}, and Solar-C \citep{Shimizu2011,Katsukawa2012,Watanabe2014}. This is partly motivated by the fact that some of the most intriguing solar phenomena, such as the heating of the solar corona, seem to have their roots in the processes that take place in the chromosphere, with the magnetic field as fundamental ingredient. For this reason, all the previously mentioned space and ground-based solar missions  have been designed with chromospheric magnetometry as a top priority. Therefore, they aim to analyse some of the key phenomena of the chromosphere such as the formation of solar spicules and their influence in the coronal heating \citep[for a recent review,][]{Tsiropoula2012}, the nature of solar fibrils \citep[for example,][]{Zirin1974,delaCruzRodriguez2011,Schad2013,Leenaarts2015}, chromospheric jets \citep{Shimizu2015}, chromospheric oscillations \citep{Leighton1962}, or magnetoacoustic waves \citep{Rosenthal2002} with simultaneous polarimetric observations of the photosphere and the chromosphere.

However, the analysis of chromospheric lines is not straightforward. Photospheric lines, like the visible Fe~{\sc i} lines, are simpler to model because it is possible to assume local thermodynamic equilibrium \citep[LTE,][]{Rutten1982} meaning that the atomic population densities can be computed from the local physical conditions at each height point. This is not feasible for chromospheric lines because they usually form in non-LTE (NLTE hereafter) conditions. Additionally, photospheric lines display larger Land\'{e} factors, with $g=1.5-3.0$, translating into higher polarimetric signals. Nevertheless, from the available chromospheric lines, the Ca~{\sc ii} infrared triplet constitutes a good candidate for the analysis of the chromospheric magnetism. The infrared lines can be modelled with a simple atomic model that greatly reduces the computation efforts. The reason is that calcium is almost entirely singly ionised under typical chromospheric conditions and non-equilibrium and partial redistribution effects are negligible for the Ca~{\sc ii} infrared lines \citep{Uitenbroek1989,Wedemeyer2011}. However, their polarimetric sensitivity is relatively low, with Land\'{e} factors close to one ($g_{8498}$=1.07, $g_{8542}$=1.10, and $g_{8662}$=0.87) and they show broad line profiles compared to photospheric lines. These properties translate into lower polarization signals than the ones produced by photospheric lines.  

In this regard, the development of spectral diagnosis methods allows us to extract the information from spectropolarimetric observations. The ultimate aim is to infer the physical conditions of the solar atmosphere that produce the observed profiles. Hereof, inversion methods provided a useful tool to achieve this goal. However, NLTE codes are not widespread because the physics of the atoms involved in the chromospheric lines is very complicated. From the available inversion codes, {\sc hazel} \citep{AsensioRamos2008}, {\sc helix}$^{\tiny +}$ \citep{Lagg2009} and {\sc nicole} \citep{SocasNavarro2015} are the most important representatives for the inversion of chromospheric lines although they are more appropriate for a few lines than for a large part of the spectrum. The first two codes are mainly designed for the inversion of He~{\sc i} 10830~\AA \ and 5876~\AA \ (or $D_3$) multiplets and take into account the joint action of Zeeman and Hanle effects. The latter is more appropriate for weak resonant lines where partial redistribution effects are not important and for atmospheric regions of moderate fields because it only considers the magnetic polarization induced by the Zeeman effect.  However, all of these codes are adequate for the lines proposed for future missions, for example, see the Solar-C SUVIT lines \citep{Katsukawa2012}. We plan to use the {\sc nicole} code to analyse the Ca~{\sc ii} 8542~\AA \ line. Our aim is to support future, ground-based and space-based, missions with detailed studies of this line. We aim to describe the polarimetric signals under different atmospheric conditions and to study the sensitivity of the line to changes of the physical parameters. Therefore, we continue the works of \cite{SocasNavarro2005,Uitenbroek2006,Pietarila2007,Cauzzi2008,delaCruzRodriguez2012,delaCruzRodriguez2013,delaCruzRodriguez2015}.

In the first part of the present paper, we focus on the analysis of the response functions to changes of the atmospheric parameters using different semi-empirical models. Then, we make use of realistic MHD simulations to infer the polarimetric signals generated by moderate fields.

\begin{figure}
\centering
\includegraphics[width=7.0cm]{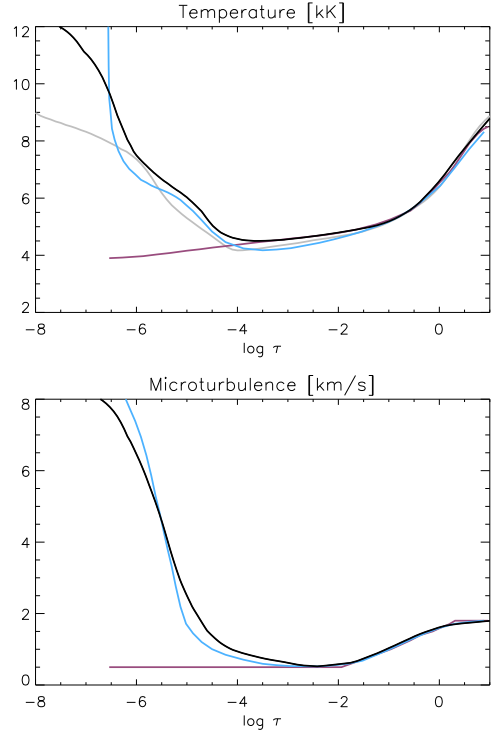}
\caption{Black colour depicts FALC model, used as the reference model in this work, while colour lines show HSRA (grey), VALC (blue) and HOLMUL (purple) models. Top and bottom panels display the temperature and microturbulence stratifications, respectively.}
\label{model}
\end{figure}

\section{Methods}

There are several codes that allow to solve the radiative transfer equation for the Ca~{\sc ii} atom, such as {\sc multi} \citep{Carlsson1986}, {\sc rh} \citep{Uitenbroek2001} and {\sc nicole} \citep{SocasNavarro2015}. The first one solves the radiative transfer equation for the intensity spectrum while the {\sc rh} and {\sc nicole} codes can compute the full Stokes vector. Moreover, from these two codes, only the latter is able to perform inversions of the Stokes profiles. Thus, {\sc nicole} is the only appropriate code for our future purposes because we aim to comprehend the information we can infer from polarimetric observations of the Ca~{\sc ii} 8542~\AA \ line as the limitations we have to bear in mind in the analysis. The reader can find the details of the {\sc nicole} code in \cite{SocasNavarro2015}, however, we summarize here some aspects of the code for completeness.

\begin{figure*}
\centering
\includegraphics[width=13cm]{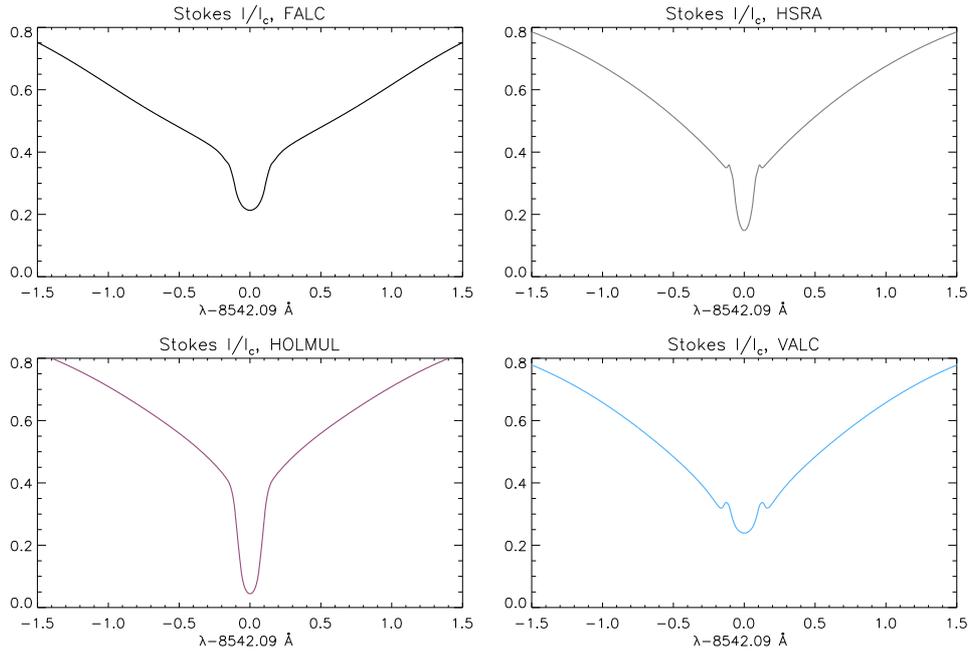}
\caption{Ca~{\sc ii} 8542~\AA \ Stokes $I$ profiles synthesized using the semi-empirical models shown in Figure \ref{model}. We followed the same colour code scheme.}
\label{stokesi}
\end{figure*}

\begin{figure}
\hspace{-0.3cm}
\includegraphics[width=7.5cm]{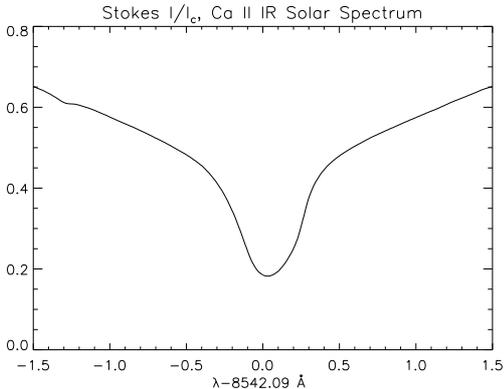}
\caption{Quiet Sun profile extracted from the BASS2000 Solar Survey Archive, based on the atlas of \citet{Atlas1973}.}
\label{atlas}
\end{figure}

NLTE atomic populations are calculated under statistical equilibrium assumption, i.e. atomic populations are computed assuming instantaneous balance between all transitions going into and out of each atomic level. The code only takes into account the polarization induced by the Zeeman effect. This means that the atomic polarization is not considered, what limits the analysis of linear polarization signals from regions with low magnetic field strengths, i.e. between 0-100~G \citep{MansoSainz2010}. Hence, the code is more suitable for moderate magnetic field regions like network patches in the quiet Sun or active regions. Additionally, the statistical equilibrium equations are solved neglecting the presence of a magnetic field, i.e. field-free NLTE atomic populations. Once the population densities are known, a full-Stokes formal solution is computed assuming that all Zeeman sub-levels are equally populated. Finally, although the code can work with 3D data cubes, each column is treated independently, and the NLTE atomic populations are solved assuming a plane-parallel atmosphere. This approximation works well in LTE conditions and for some strong NLTE lines where the 3D radiation field does not play an important role for the computed population densities. In the case of the Ca~{\sc ii} infrared lines, this assumption is fairly accurate \citep{delaCruzRodriguez2012}.

Regarding the Ca~{\sc ii} atomic model, {\sc nicole} considers five bound levels plus a continuum (see Figure 1 of \cite{Shine1974} for a representation of the model). In addition, the code assumes complete frequency redistribution that is valid for weak resonant lines and subordinate lines, as for the Ca~{\sc ii} 8542~\AA \ line \citep{Uitenbroek1989}. As additional note, {\sc rh} is the only code that takes into account partial redistribution effects to compute the full Stokes vector, what makes it more suitable for studying the polarization properties of strong resonant lines as  Ca~{\sc ii h} \& {\sc k} or Mg~{\sc ii} h \& k lines.

Of the three spectral lines included in the Ca~{\sc ii} infrared triplet, the Ca~{\sc ii} 8542.09~\AA \ line is the one that shows the largest Land\'{e} factor. Additionally, it does not show blends with other solar spectral lines, as happens with  the Ca~{\sc ii} 8498~\AA \ line. For these reasons, this line has been extensively used in previous works as in the present one. Therefore, in our study, we make use of {\sc nicole} and we focus on the analysis of the spectral properties of the Ca~{\sc ii} 8542.09~\AA \ line from NLTE synthesis. We are going to use semi-empirical models as FALC \citep{Fontenla1993}, HSRA \citep{Gingerich1971}, VALC \citep{Vernazza1981}, and HOLMUL \citep{Holweger1974}, see Figure \ref{model}. Although there are some distinct aspects between these models, as the hydrogen and electron densities, the most noticeable differences are the temperature and microturbulence stratifications. These differences are translated into dissimilar Stokes $I$ profile shapes, Figure \ref{stokesi}. The HOLMUL model generates a very deep profile due to the low temperatures at higher layers, while the HSRA and the VALC models produce abrupt jumps at the wavelengths pertaining to the transition from the wings to the line core, probably due to a lack of microturbulence at higher layers and the different temperature stratification (see Figure \ref{model}). However, the model that produces a Stokes $I$ profile closest to the observed quiet Sun intensity spectrum (Figure \ref{atlas}) is the FALC model. For this reason, we used this atmospheric model as reference for following studies. We synthesized all the profiles showed in Figure \ref{stokesi} assuming a null macroturbulence producing narrow Stokes $I$ wings in comparison with the atlas profile. On the other hand, we considered the original microturbulence of each model what generates line core widths similar to the one showed by the atlas profile (see Figure \ref{atlas}). In addition, the line of sight (LOS, hereafter) velocity is also null for the four models, therefore without gradients, generating a symmetric Stokes $I$ line core at rest.

\section{Response Functions}

Response functions (RF) provide information about the sensitivity of a given spectral line to perturbations of the physical parameters of the atmospheric model \citep{Landi1977}. The concept of RF arises from the analysis of the radiative transfer equation under first order perturbations in the physical parameters. A perturbation $\delta x(\tau$\footnote{The parameter $\tau$ refers to the optical depth evaluated at a wavelength where there are no spectral lines, i.e. continuum. In our case, this wavelength is 5000~\AA.}) in a single physical magnitude, $x(\tau)$, will produce a modification $\delta \textbf{I}$ of the emergent Stokes spectrum, $\textbf{I}$. Following  \cite{Landi1977} with the numerical notation of \cite{RuizCobo1992} we can write this statement in a differential equation for $\delta \textbf{I}$ whose formal solution is

\begin{equation}
\delta \textbf{I}(\lambda) = \int_{0}^\infty \textbf{R}(\lambda,\tau) \delta x d\tau 
\label{rfeq}
\end{equation}
where $\textbf{R}(\lambda,\tau)$ is the response function vector that depends on the wavelength of the spectral line and the optical depth of the perturbed layer in the atmospheric model. However, although this expression is the one that usually appears in the literature, the integration is often done in $\log \tau$ steps instead of $\tau$. For this reason, we are going to rewrite equation \ref{rfeq} as function of $\log \tau$. Thus,

\begin{equation}
\delta \textbf{I}(\lambda) = \int_{\log \tau_1}^{\log \tau_2} \hat{\textbf{R}}(\lambda,\tau) \delta x d(\log \tau) 
\label{rfeq2}
\end{equation} 
where $\hat{\textbf{R}}(\lambda,\tau)=\ln(10)\tau\textbf{R}(\lambda,\tau)$. The lower and upper limits, i.e. $\tau_1$ and $\tau_2$, are numbers much smaller and much larger than 1, respectively.   

The response function also depends on the heliocentric angle and on the atmospheric model itself \citep{Landi1977}, indicating that the results presented in this section depend on the model employed to obtain them and cannot be interpreted as absolute findings. Moreover, there are four $\hat{\textbf{R}}(\lambda,\tau)$ for a single set of the previous parameters, one for each Stokes parameter. Then, calculating the $\hat{\textbf{R}}(\lambda,\tau)$, we are going to be able to analyse in detail the response of a given Stokes profile to changes of the atmospheric physical parameters at a given optical depth.

\begin{figure*}
\centering
\includegraphics[width=16.5cm]{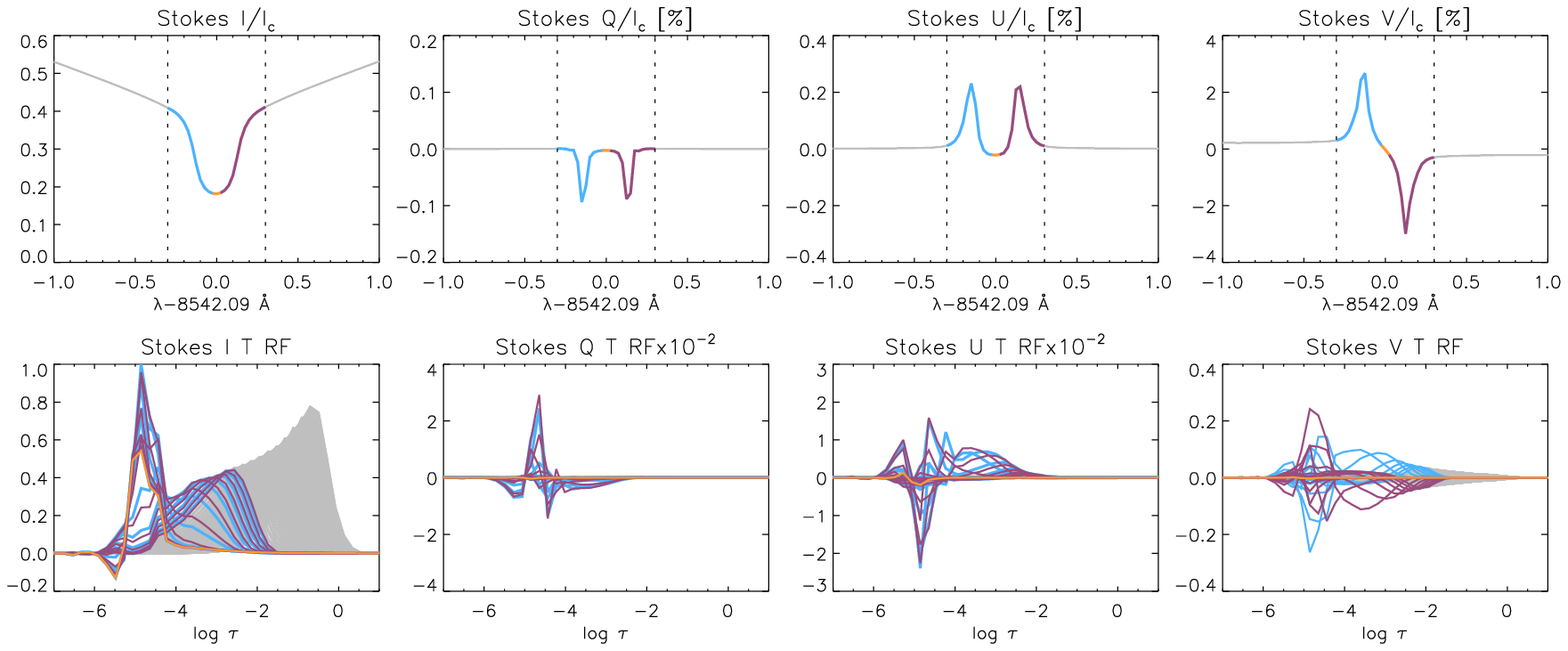}
\includegraphics[width=16.5cm]{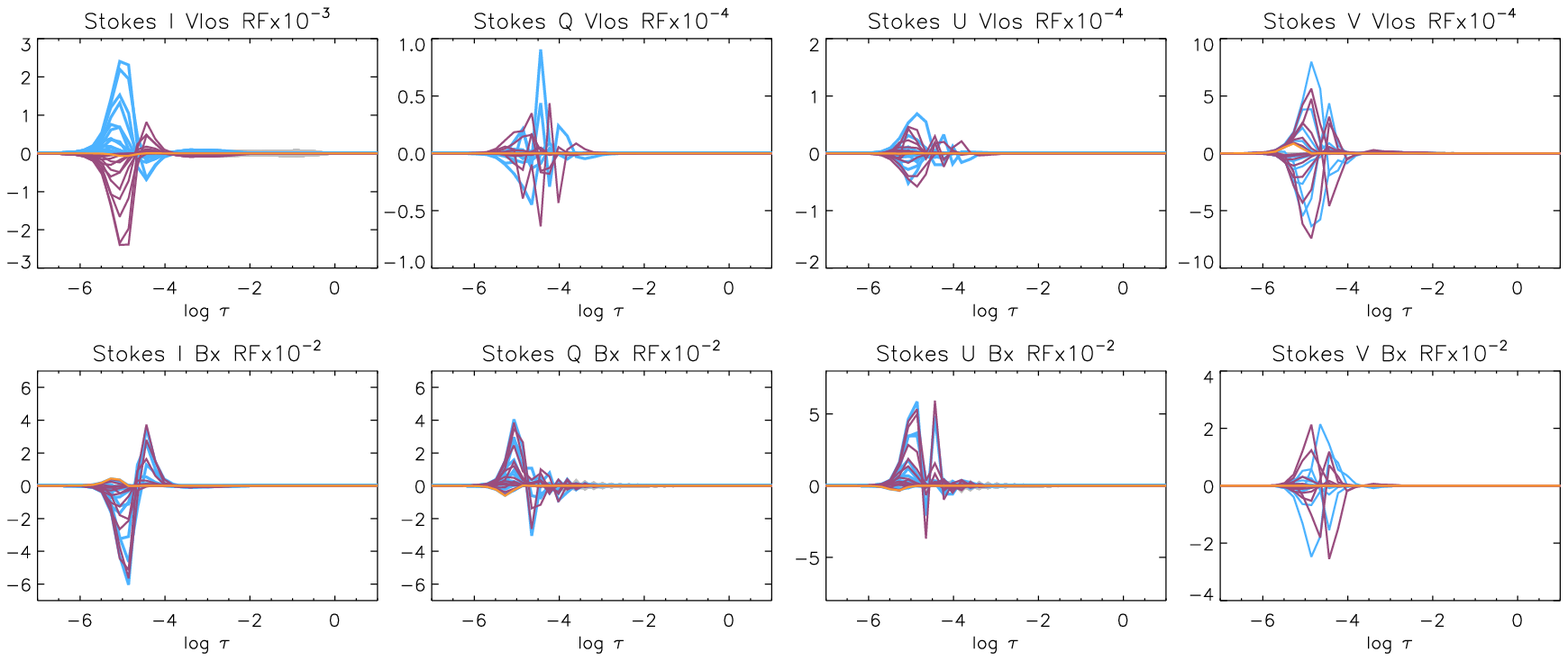}
\includegraphics[width=16.5cm]{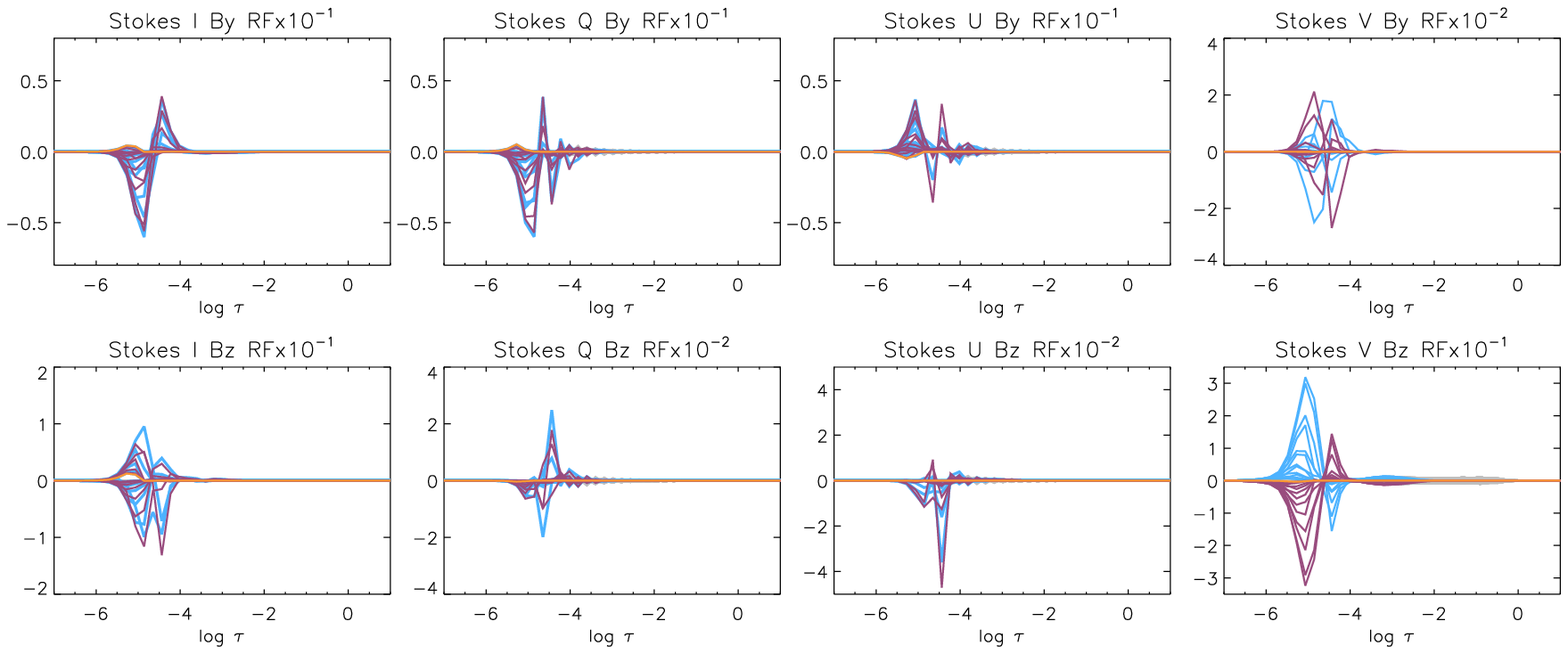}
\caption{Synthetic Stokes profiles and response functions obtained from the FALC model with a magnetic field strength of 500 G and a magnetic field inclination and azimuth of 45 degrees. We display the RF versus optical depth in 1D plots including all wavelengths. Colours designate the different spectral regions depicted in the Stokes profiles of first row and every RF is normalised to the maximum of the Stokes $I$ RF to the temperature. Large positive/negative values indicate that the spectral line is sensitive to perturbations of a given magnitude at a given layer.}
\label{rfFig}
\end{figure*}

\begin{table}
  \caption{Magnitudes of the perturbation $\delta x$ used to compute the response functions.}
  \label{tableRF}
  \begin{tabular}{lcc} 
	\hline
	Variable              & Value & Units  \\
	\hline
	$\delta$T             &  1    &   K   \\
	$\delta V_{\rm los}$  &  10   &   m/s \\
	$\delta B_x$          &  25   &   G    \\
	$\delta B_y$          &  25   &   G    \\
	$\delta B_z$          &  25   &   G    \\ 
	\hline
  \end{tabular}
\end{table}

\begin{figure*}
\centering
\includegraphics[width=14cm]{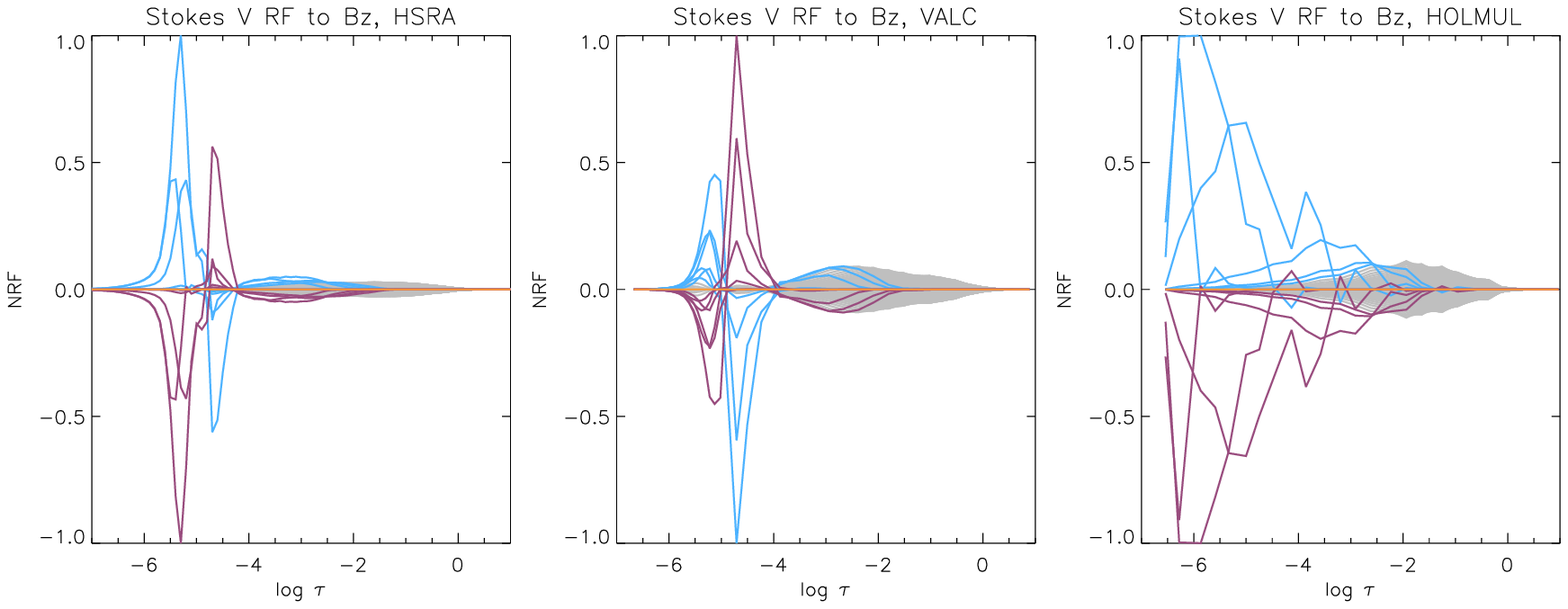}
\caption{Stokes $V$ RF to $B_z$ computed with the semi-empirical models showed in Figure \ref{model}. We normalized each RF to its maximum value. In addition, we followed the same coloured scheme used in Figure \ref{rfFig}.}
\label{rfmodel}
\end{figure*}

There are several ways to calculate RF. From analytical cases (see \cite{RuizCobo1992} and references within) that start from the assumption of an atmosphere in LTE, to numerical approaches where the line of study is no longer in LTE \citep{Fossum2005,Uitenbroek2006}. In our case, as we plan to study the Ca~{\sc ii} 8542 \AA, we cannot assume LTE conditions. We opted to estimate Eq. \ref{rfeq2} using numerical perturbations to the physical parameters of the model atmosphere \citep[e.g., ][]{SocasNavarro2004b}. We selected a reference model $M$ that we built from the FALC model, black colour in Figure \ref{model}, including a constant magnetic field of 500 G and 45 degrees of inclination and azimuth. This configuration guarantees that none of the Stokes profiles is zero. Then, we calculated the Stokes profiles produced by a perturbation $\delta x(\tau)$ in a given physical parameter. We created a model $M^{+}$ that is identical to the model $M$ but, in a single optical depth position, the value of the physical parameter is the original value plus $\delta x$, in our study the perturbation is not optical depth dependent. At the same time, we created a model $M^{-}$ that is exactly the same as $M^{+}$ but instead of having the original value plus $\delta x$, it has the original value minus $\delta x$. Therefore, we designed two models that in a single optical depth they are perturbed by $\pm\delta x$. The result will be two Stokes profiles $I_{i}^{+}(\lambda)$ and $I_{i}^{-}(\lambda)$ (we focus just on Stokes $I$ for simplicity but the argument is valid for the rest of the Stokes parameters) obtained through a perturbation at an optical depth $\tau _i$ (the $\delta x$ values used in this study are indicated in Table \ref{tableRF}). The greater the difference between these two Stokes profiles the larger the sensitivity of the Stokes parameter to the perturbation $\delta x$. The next step consists in repeating the synthesis for different optical depths, obtaining for every case a set of Stokes profiles $I_{i}^{+}(\lambda)$ and $I_{i}^{-}(\lambda)$, where $i$ stands for the optical depth. Then, we can calculate the numerical RF to a perturbation on the physical parameter $x$ as

\begin{equation}
\hat{\textbf{R}}_x(\lambda,\tau)=\frac{1}{C } \left(
\begin{array}{cccc}
d(\tau_0,\lambda_0)    & d(\tau_0,\lambda_1)  & \ldots & d(\tau_0,\lambda_j) \\
d(\tau_1,\lambda_0)    & d(\tau_1,\lambda_1)  & \ldots & d(\tau_1,\lambda_j) \\
\vdots & \vdots & \vdots & \vdots \\
d(\tau_i,\lambda_0)    & d(\tau_i,\lambda_1)  & \ldots & d(\tau_i,\lambda_j) \\
\end{array} \right)
\end{equation}
with $d(\tau_i,\lambda_j)=I_{i}^{+}(\lambda_j)-I_{i}^{-}(\lambda_j)$ and $C=2\delta x\delta (\log\tau)$, $\delta (\log\tau)$ being constant for all heights (we interpolated the original FALC model to an uniform grid of optical depths) and equal to 0.2. Optical depth values extend from the bottom to the top of the FALC model, i.e. $\log \ \tau=[1,-8]$, and the wavelength values cover $\lambda$=[$\lambda_c$-3.5,$\lambda_c$+3.5]~\AA,  $\lambda_c$ being the line centre wavelength. This method provides an easy way to compute the sensitivity of any spectral line to changes in the different physical parameters and can be applied to any atmospheric model.

We computed the RF ($\hat{\textbf{R}}_x(\lambda,\tau)$) for temperature, LOS velocity and the three components of the magnetic field defined as follows,

\begin{align}
B_x &= B\sin\theta\cos\phi \\
B_y &= B\sin\theta\sin\phi \\
B_z &= B\cos\theta 
\end{align}
where $B$ stands for the magnetic field strength while $\theta$ and $\phi$ are the inclination and azimuth angles, respectively. The resulting RF, i.e. $\hat{\textbf{R}}_x(\lambda,\tau)$, are plotted in Figure \ref{rfFig}. Each column of this figure provides information about a Stokes parameter, $I$, $Q$, $U$, and $V$. We plot in the first row the synthetic Stokes profiles obtained from the chosen model $M$. We display with different colours line wings (grey), blue core (blue), red core (purple), and line core centre (orange) wavelengths. We tentatively separate these wavelength regions with colours because they correspond to different optical depth values, i.e. different atmospheric heights. In addition, we reduced the spectral range in the plot to $\pm 1$~\AA \ to improve the line core visibility. The following rows of Figure \ref{rfFig} represent the RF for temperature, LOS velocity and the three components of the magnetic field. Although the usual way to represent RF is a 2D plot (see, for instance, \cite{SocasNavarro2004b} or \cite{Uitenbroek2006}) we opted first for a different approach, similar to the one used in \cite{RuizCobo1992}, for discussing the response functions, although we will include later the traditional 2D plot. We display 1D plots where we can clearly examine the RF at different optical depths. But, instead of plotting one line corresponding to a wavelength of interest \citep[see Figure 1 of][]{RuizCobo1992}, we included all the wavelengths synthesized. In that sense, we can see the global response to each physical parameter as we examine different spectral regions, i.e. different colours. In addition, we normalized every RF to the maximum of the Stokes $I$ RF to the temperature, the largest RF for our model. As an additional comment, we keep the sign of the RF to study their variation with height.

\begin{figure*}
\hspace{+0.5cm}
\includegraphics[width=17.5cm]{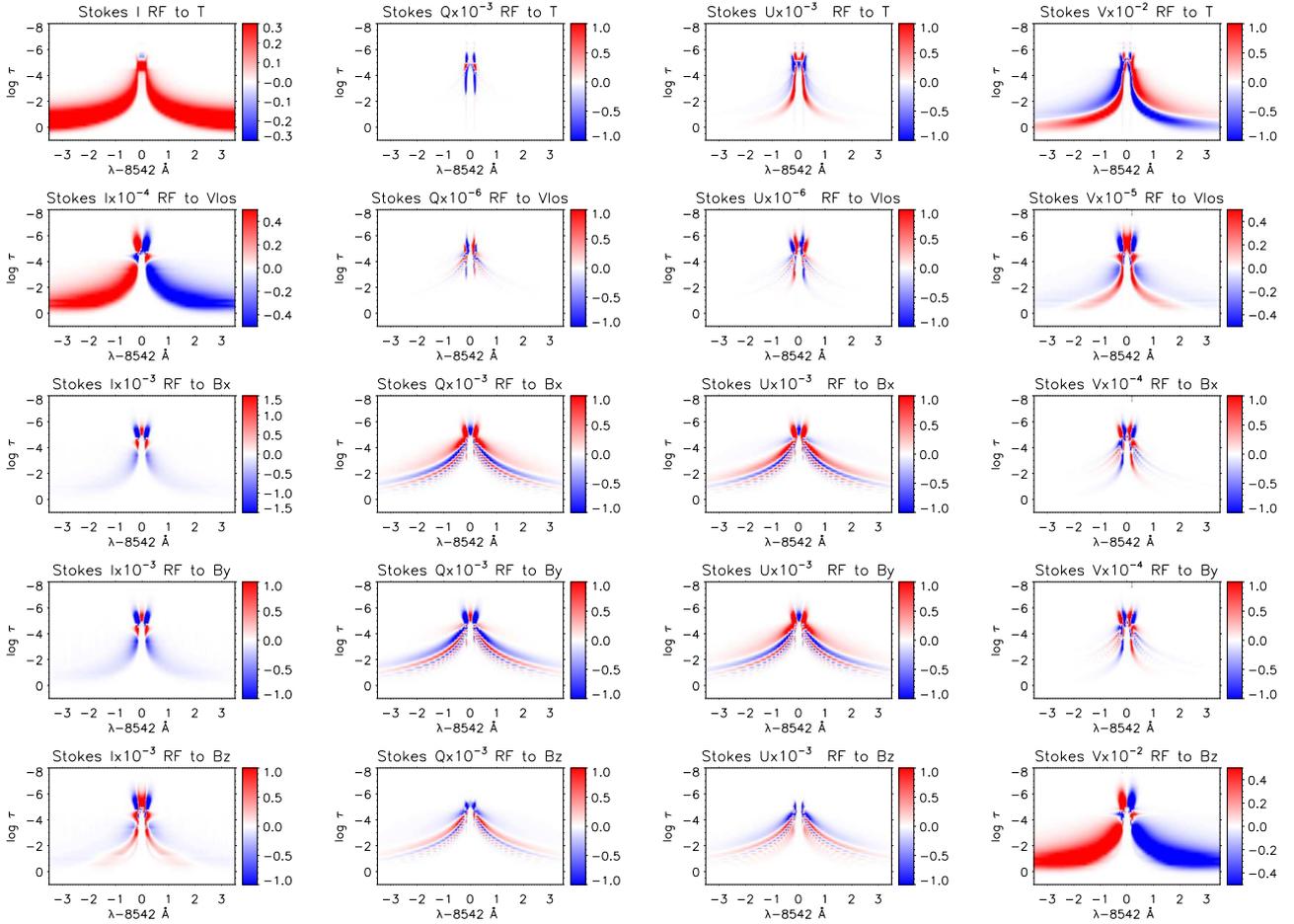}
\caption{Two-dimensional representation of the RF shown in Figure \ref{rfFig}. White areas indicate non-sensitive regions while red or blue colour depicts the locations where the spectral line is sensitive to changes on the atmospheric parameters. In addition, every RF is normalised to the maximum of the Stokes $I$ RF to the temperature.}
\label{rf2d}
\end{figure*}

The way to interpret the results is the following: large positive/negative values indicate that the spectral line is sensitive to perturbations of a given magnitude at a given layer, while low or null values indicate that the uncertainty in this region is very large, i.e. we cannot retrieve reliable information of the physical parameters in this region. Focusing on the second row, the RF to temperature perturbations, we can see large grey values at low heights in Stokes $I$ meaning that the Stokes $I$ profile wings are sensitive to temperature changes in those regions. Then we have blue/purple and orange lines in the upper photosphere with a strong peak around $\log$~$\tau=-5$. Thus, the line core is sensitive to temperature changes that take place in the lower-middle chromosphere. Then the RF falls to zero very quickly above this layer as well as below $\log$~$\tau=0$. Remarkably, there is also a strong decrease of the RF around $\log$~$\tau=-4$ that indicates that the line is not very sensitive to temperature changes in this region. Thus, we will not be able to infer accurate information from these layers. If we move to Stokes $Q$, and $U$ RF for temperature we can see that the linear polarization parameters are not very sensitive to changes in temperature showing very small values, although the shape of the RF slightly resembles the Stokes $I$ one at upper layers. For Stokes $V$, we found larger values with grey lines in the bottom of the photosphere and blue/purple lines in upper layers. We can also note that the Stokes $V$ RF is anti-symmetric (compare blue and purple colour above $\log$~$\tau=-2$), something that we expect from the model we used because there are no gradients in the line of sight velocity nor in the magnetic field stratifications.

The third row shows the results for the LOS velocity RF. Starting with the Stokes $I$ profile we can see that the RF of the spectral line wings (grey) is very low in the bottom part of the photosphere, indicating that the line is not very sensitive to LOS velocity changes in those regions. This property has been discussed in \cite{delaCruzRodriguez2015} where they explain that this behaviour is due to the Lorentzian contribution that broadens the line wings, masking the presence of steep intensity gradients. However, as we move up in the atmosphere, until $\log$~$\tau=-4$, we can find an increase of sensitivity from the line wings and part of the line core (blue/purple). We encounter again a low response around $\log$~$\tau=-4$, although less pronounced than in the temperature RF, and a strong increase of sensitivity above this layer. In this case, the sensitivity at upper layers goes beyond and almost reaches $\log$~$\tau=-6$. For the rest of the Stokes parameters, we found similar results although with less sensitivity.

The results for the horizontal components of the magnetic field $B_x$ and $B_y$ are shown in the fourth and fifth rows. They behave in a very similar way as the Stokes $Q$ and $U$ RF showing sensitivities above $\log$~$\tau=-1$ until $\log$~$\tau=-5.5$. However, in this case, the RF shape is much more complex than what we saw for the previous atmospheric parameters. Regarding Stokes $I$ and $V$ RF, we only find sensitivity at upper layers in the line core wavelengths.

\begin{figure*}
\centering
\includegraphics[width=17.3cm]{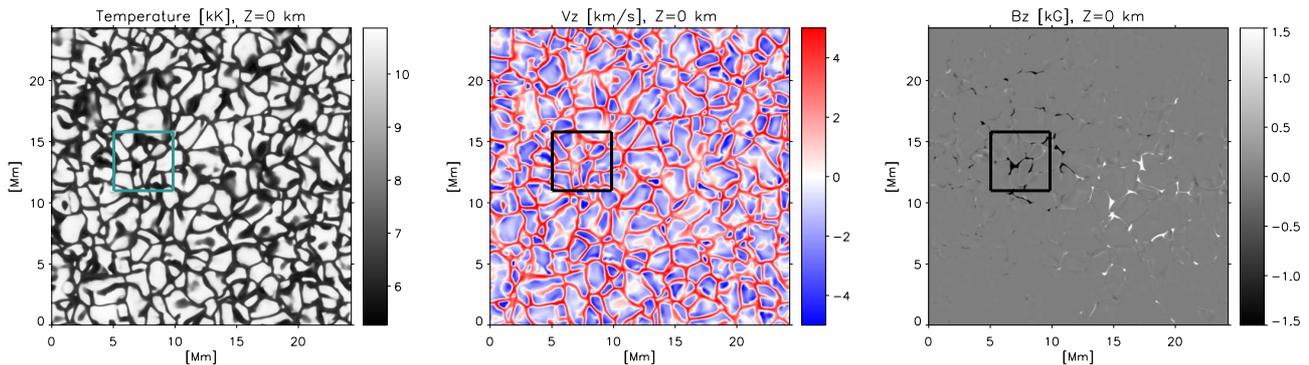}
\caption{Simulation snapshot used in this work. From left to right, temperature, longitudinal velocity, and longitudinal magnetic field. We show the spatial distribution of these physical parameters at the height Z=0 km. The square box marks the region we are going to analyse in detail in this work.}
\label{context}
\end{figure*}

Last row of this figure shows the RF to the longitudinal field $B_z$. In this case, we can see that Stokes $V$ RF is the dominant and it provides sensitivity from the bottom of the photosphere, from $\log$~$\tau=0$, to the middle chromosphere, around  $\log$~$\tau=-5.5$. It is clear the height separation between grey and blue/purple colours. Additionally, we can also see a change of sign in the RF for the line core wavelengths at the optical depth of $\log$~$\tau=-4$ and $-4.7$. This change of sign indicates that, when the RF is positive, increasing $B_z$ produces larger Stokes $V$ amplitudes while, when the RF is negative, increasing $B_z$ produces smaller Stokes $V$ amplitudes. This is something that we would expect in the strong field regime. However, we found the same results using a magnetic field with lower strength, i.e. $B$=50 G (not shown in this work). Therefore, we believe that, in this case, the change of sign is related to the change of the temperature gradient of the FALC model that induces a change in the source function gradient \citep[see][]{Uitenbroek2006}. We compared these results with different models, i.e. HSRA, VALC and HOLMUL models (see colour lines in Figure \ref{model}) and we find a different behaviour only for the case of the HOLMUL model (see Figure \ref{rfmodel}). The other two semi-empirical models, first and second columns of Figure \ref{rfmodel}, show a similar RF shape although with small differences; for example, the VALC RF shows less sensitivity at $\log$~$\tau\sim-4.7$ than at $\log$~$\tau\sim-5.5$, opposite to what we found for FALC and HSRA. Additionally, the HOLMUL RF shows artificial large sensitivities in the top of the atmosphere due to the wrong LTE temperature at these layers. Hence, for the configuration of atmospheric parameters used in this study, the change of sign on the Stokes $V$ RF to $B_z$ is related to the change of the temperature gradient. Remarkably, other models with more complicated magnetic field stratifications, as jumps along the line of sight, could produce a change in the Stokes $V$ RF to the magnetic field, see Figure 1 of \cite{RuizCobo1992}. Additionally, we want to emphasize that RF are model dependent, so these results should be taken as illustrative instead of absolute findings.

Finally, we present in Figure \ref{rf2d} the 2D traditional representation of RF for completeness. We want to emphasize through this figure the possibility to extract unambiguous information from the Stokes parameter. For example, if we focus on the bottom row, rightmost panel, we can see that the Stokes $V$ profile is sensitive to changes in the magnetic field at different heights but also at different wavelengths. Therefore, a change in the Stokes profile in a single wavelength would be attributed to a change in the longitudinal field at a single height. For comparison, the $H_\alpha$ line displays the opposite behaviour, see Figure 1 of \cite{SocasNavarro2004b}, showing strong sensitivity on the core wavelengths for changes at low photospheric heights as well as low chromospheric heights, which makes it complicated to infer the height stratification of the longitudinal magnetic field. However, we still plan to perform more studies in the future performing inversions of the Stokes profiles using different longitudinal magnetic fields to check if a change of the Stokes profiles in a single wavelength can be unambiguously attributed to a perturbation at a single height.

\begin{figure*}
\centering
\includegraphics[width=17.3cm]{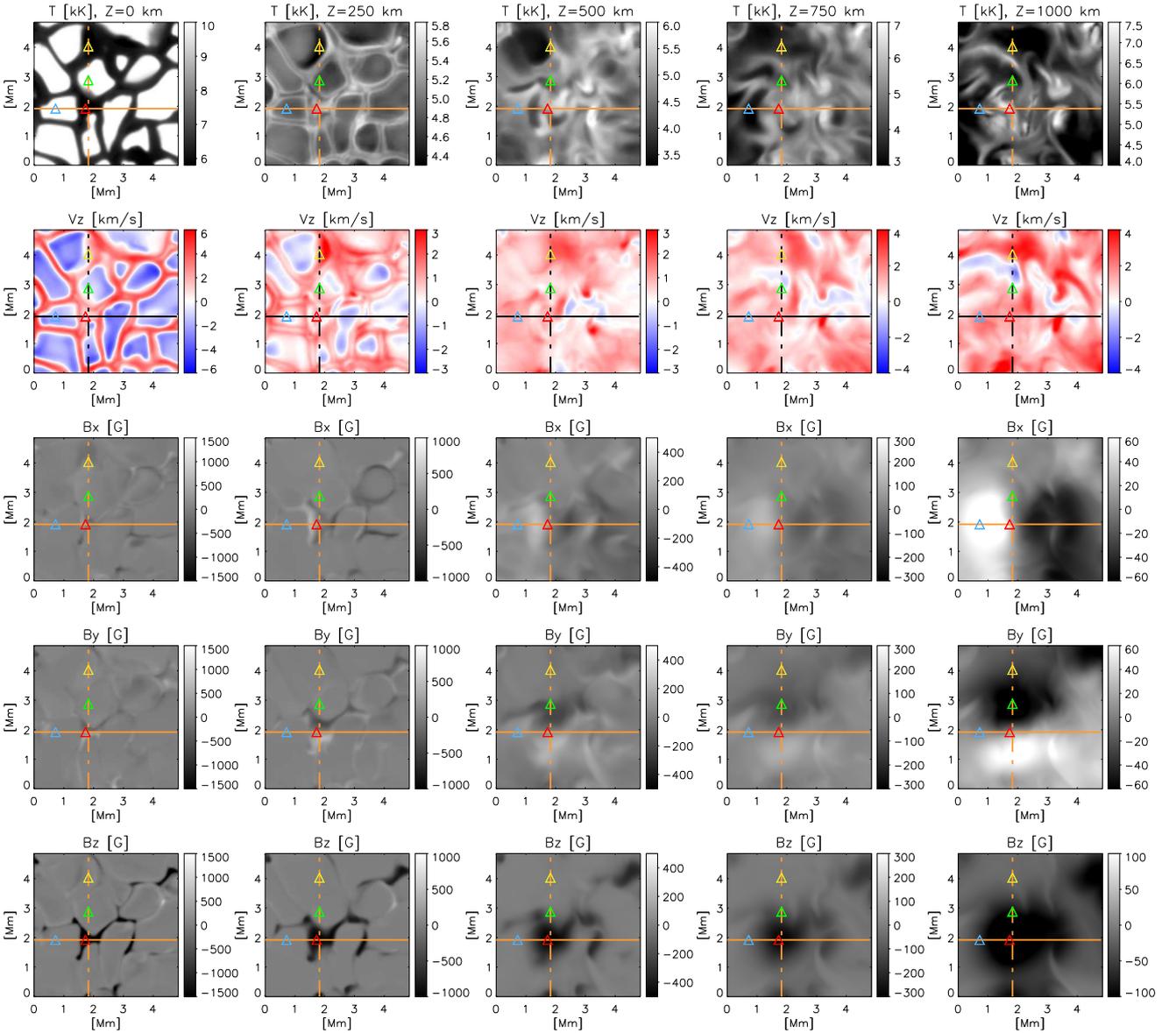}
\caption{Spatial distribution of the atmospheric parameters at different heights. Each column corresponds, from left to right, to the height values Z=[0, 250, 500, 750, 1000] km while each row displays, from top to bottom, the temperature, LOS velocity and the three Cartesian components of the magnetic field vector. Orange lines designate the regions we are going to examine later through a vertical cut of the atmosphere. Triangles mark the location of selected pixels we analyse in detail in following sections.}
\label{r1}
\end{figure*}

\section{Simulations}\label{sims}

We  employed  a  snapshot  from  a  3D  MHD  simulation  calculated with  the {\sc bifrost} code  \citep{Gudiksen2011}. The simulation includes a patch of enhanced network \citep{Carlsson2016} that is publicly available as part of the IRIS mission \citep{dePontieu2014}. The horizontal size is $24\times24$~Mm$^{2}$, while its vertical dimension encompasses the upper convection zone, photosphere, chromosphere, and corona, having a total size of 16.8~Mm. The total vertical range goes from 2.4~Mm below to 14.4~Mm above average optical depth unity at $\lambda$=5000~\AA. The horizontal grid spacing is constant with a value of 48~km and the vertical direction spacing is non-equidistant  with  a  spacing  of  19~km between \textit{Z}=-1 and 5 Mm, and increasing toward the upper and lower boundaries. The  simulation  includes optically  thick  radiative transfer considering scattering in the photosphere and low chromosphere, parametrized NLTE radiative losses in the upper chromosphere,  transition  region  and  corona,  thermal  conduction along magnetic field lines, and an equation of state that  includes  the effects  of  non-equilibrium  ionization  of  hydrogen; see \cite{Gudiksen2011,Carlsson2012} for more details. We show in Figure \ref{context} the spatial distribution at \textit{Z}=0~km of temperature, longitudinal velocity, and longitudinal magnetic field. This snapshot simulates an \textit{enhanced network} region \citep{Carlsson2016} dominated by two magnetic patches of opposite polarity separated by 8~Mm (rightmost panel) and corresponds to the simulation time t~$=3850$~s. The granulation pattern is clearly visible in the temperature and velocity panels. Therefore, this snapshot represents a low magnetised solar atmosphere. Additionally, there are some regions of special interest as chromospheric bright points or very cool chromospheric regions located over relatively large granules. However, although it is worth to examine those regions, we decided to leave them for another work and focused the present paper on the magnetic flux tube enclosed inside the square.

\begin{figure*}
\includegraphics[width=17.7cm]{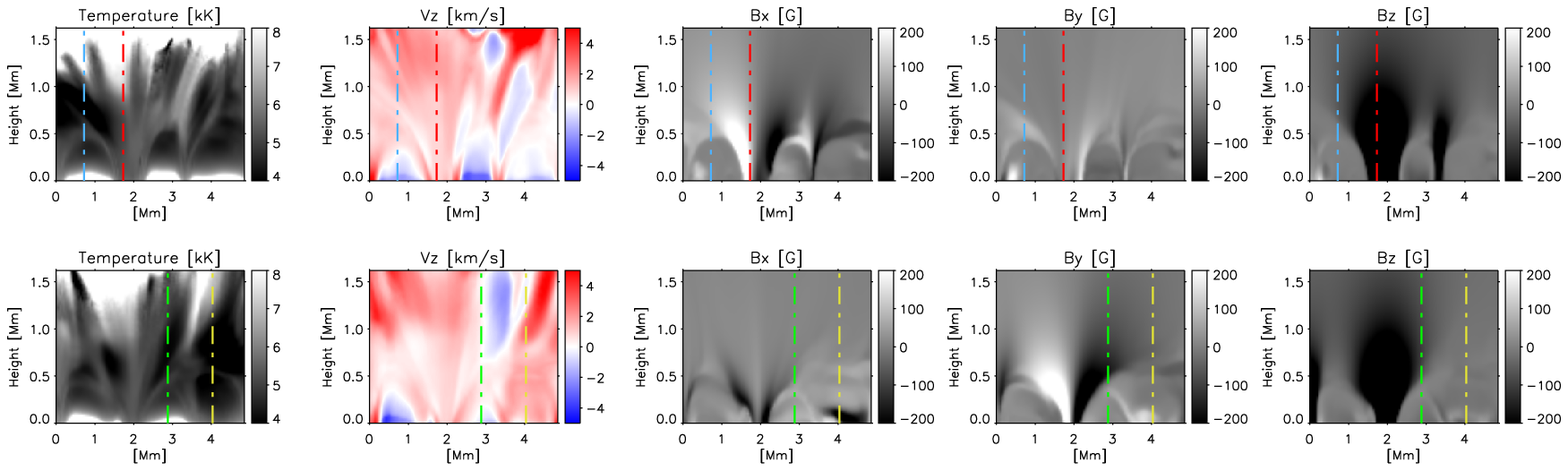}
\caption{Vertical cut of the temperature (leftmost column), longitudinal velocity (second column), and magnetic field vector (rightmost columns). Top row corresponds to the solid orange line in Figure \ref{r1} while second row displays the information contain in the pixels that belong to the dashed orange line of Figure \ref{r1}. Coloured lines designate the location of the triangles of Figure \ref{r1}.}
\label{vertical}
\end{figure*}

\subsection{Spatial distribution at different heights}

We are going to study in this section the spatial distribution of temperature, longitudinal velocity, and the three components of the magnetic field at different heights, see Figure \ref{r1}. In that sense, we are going to continue the work of \cite{delaCruzRodriguez2013}. In addition, we will also study the polarimetric signals we can expect from this network region in following sections.

Starting with the temperature panels, first row, we can see the granulation pattern, with hot granules and cool intergranular lanes, in the leftmost part of the figure. The temperature contrast between structures is high producing a sharp and crispy image. Second column displays the opposite behaviour showing cool granules and hot intergranules, the so-called \textit{reverse granulation} pattern that was first observed using Ca~{\sc h \& k} lines \citep{Evans1972}. If we examine higher layers, three last columns, we can find the highly dynamic pattern that is usually observed in chromospheric observations \cite[for instance, ][]{Rutten2007} where thin structures twist and bend with spatial sizes smaller than granular features. Concerning the longitudinal velocity (second row), we can distinguish the granulation pattern at the bottom of the photosphere (Z=0 km), with granules slowly moving upward (blue) and intergranules moving downward with larger velocities (red). Both structures are joint by narrow white locations where the longitudinal velocity is null. These white regions delimit the change of sign of the longitudinal velocity from the granular structure to the intergranular lane and vice versa. Second column shows a decrease of the longitudinal velocity in the granulation structures, although there is no opposite landscape as happen with the temperature. Interestingly, intergranular lanes show a white narrow line in their centre, pointing to a null velocity at this height. For higher layers, the velocity pattern is similar to the temperature one, with complex and thin structures, relatively low velocities (about $\pm$3~km/s) where downflows are dominant. 

Regarding the magnetic field vector, we can see that its horizontal components, third and fourth rows, display a similar behaviour. Magnetic field signals are low at the bottom of the photosphere and they become more important as we examine higher atmospheric layers, see two rightmost panels. The reason behind this behaviour is that the magnetic field lines are mainly vertical at the low photosphere, at the base of the flux tube, and they get inclined at larger heights due to the expansion of the magnetic field as a consequence of the gas pressure reduction. Finally, the longitudinal magnetic field (bottom row) is predominately unipolar at all heights, being narrow and strong, above kG values, at the low photosphere (leftmost panels) while it fills almost all the selected area with weak fields at higher layers (rightmost panel). We want to remark that the magnetic field strength drops from 1-2 kG at \textit{Z}=0 km to $\sim200$ G at \textit{Z}=1000 km, thus, the longitudinal magnetic field strength is relatively low in the chromosphere, even in the network patch.

\subsection{Vertical stratification}

We show in Figure \ref{vertical} two-dimensional cuts that cross the network patch from two different views (see orange lines in Figure \ref{r1}) to examine their vertical stratification. If we focus on the upper row,  we can see that the temperature stratification, leftmost panel of Figure \ref{vertical}, shows the granulation pattern at lower layers and then, it evolves displaying a shape similar to the water of a fountain that expands with height. The temperature distribution in this region displays uniform and relatively hot values (grey areas) that are also present in the longitudinal velocity pattern (second panel) where a gentle downflow fills the same area. Third and fourth columns display the vertical distribution of the horizontal component of the magnetic field, showing signals in the edges of the flux tube at the upper photosphere and chromosphere. Finally, in the rightmost panel, we clearly see the vertical distribution of two magnetic flux tubes, thin at the bottom of the photosphere and thicker at the top of the atmosphere due to the expansion with height of the magnetic field. 

Bottom row of Figure \ref{vertical} shows a similar behaviour for all the physical parameters although, in this case, as we are looking to the flux tube from a different point of view, the horizontal component of the magnetic field shows an opposite distribution. In addition, comparing both perspectives, we can see that the magnetic structure is not symmetric.
 
\subsection{Selected pixels}\label{pixelsofinterest}

\begin{figure*}
\includegraphics[width=17.5cm]{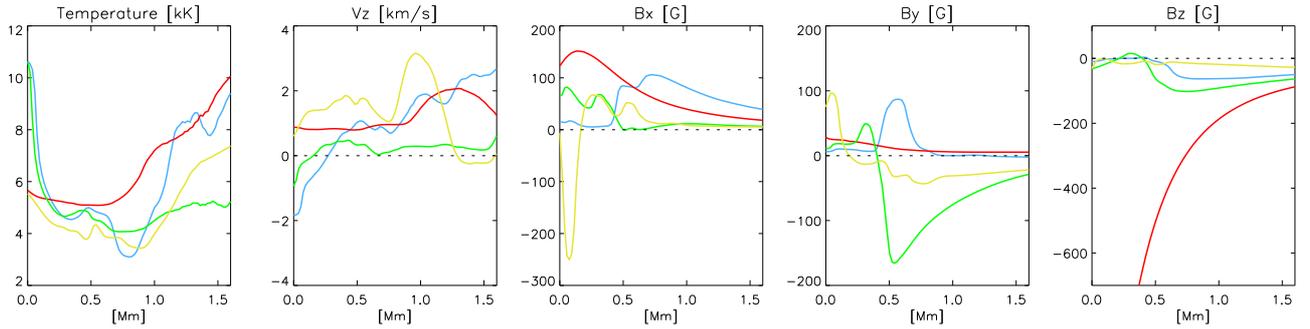}
\vspace{-0.3cm}
\caption{Atmospheric stratification of the pixels marked with coloured triangles in Figure \ref{r1} and coloured lines in Figure \ref{vertical}. The panel distribution is the same as in the mentioned figures. We also used the same colour scheme for differentiating the selected pixels.}
\label{uni}
\end{figure*}

We indicate with coloured triangles in Figure \ref{r1} and coloured lines in Figure \ref{vertical} the location of selected pixels that we are going to study in detail in this section. We opted to select regions that will probably generate peculiar profiles in the synthesis process. In this regard, we select the centre of the magnetic flux tube (red), two canopy regions (blue and green) and a region with low magnetic field (yellow). We plot the atmospheric parameters of these pixels versus geometrical height on Fig. \ref{uni}.

\begin{figure*}
\hspace{-0.9cm}
\includegraphics[width=17.5cm]{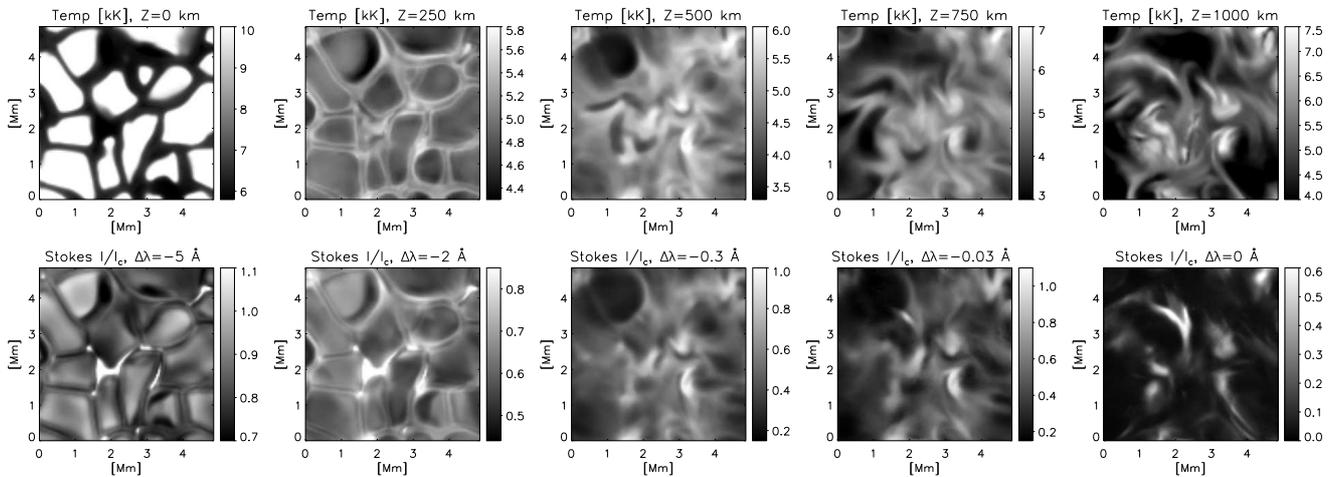}
\caption{Temperature spatial distribution at different heights (top row) and Stokes $I$ intensity at different monochromatic wavelengths (bottom row).}
\label{sivar}
\end{figure*}

The temperature stratification, leftmost panel, for the blue, green and yellow pixels show a vertical distribution that resembles a quiet Sun semi-empirical model (see Figure \ref{model}). In the case of the red pixel, we found that the temperature is higher and flatter than the other models. This is due to the strong magnetic field contained in the centre of the flux tube  \cite[see also the discussion of][]{delaCruzRodriguez2013}. The longitudinal velocity is similar for all pixels with low amplitude variations with height and showing mainly redshifted velocities. Moving to the horizontal components of the magnetic field we found that blue and green pixels show almost zero field strength below 500 km and strong magnetic fields above these layers. Yellow pixels shows very low magnetic fields values for all heights except at the very low atmosphere where probably the line is not very sensitive. Finally, the longitudinal field panel shows an almost null value for the yellow pixel, and the same canopy structure for the green and blue pixels. In the case of the red pixel (flux tube centre), we found a very large magnetic field with 1.6 kG strength at $Z$=0 km that monotonically decreases to 100 G at 1500 km. We use a reduced vertical axis in order to visualize the jump in the line of sight magnetic field of the green and blue pixels.

\begin{figure*}
\includegraphics[width=17cm]{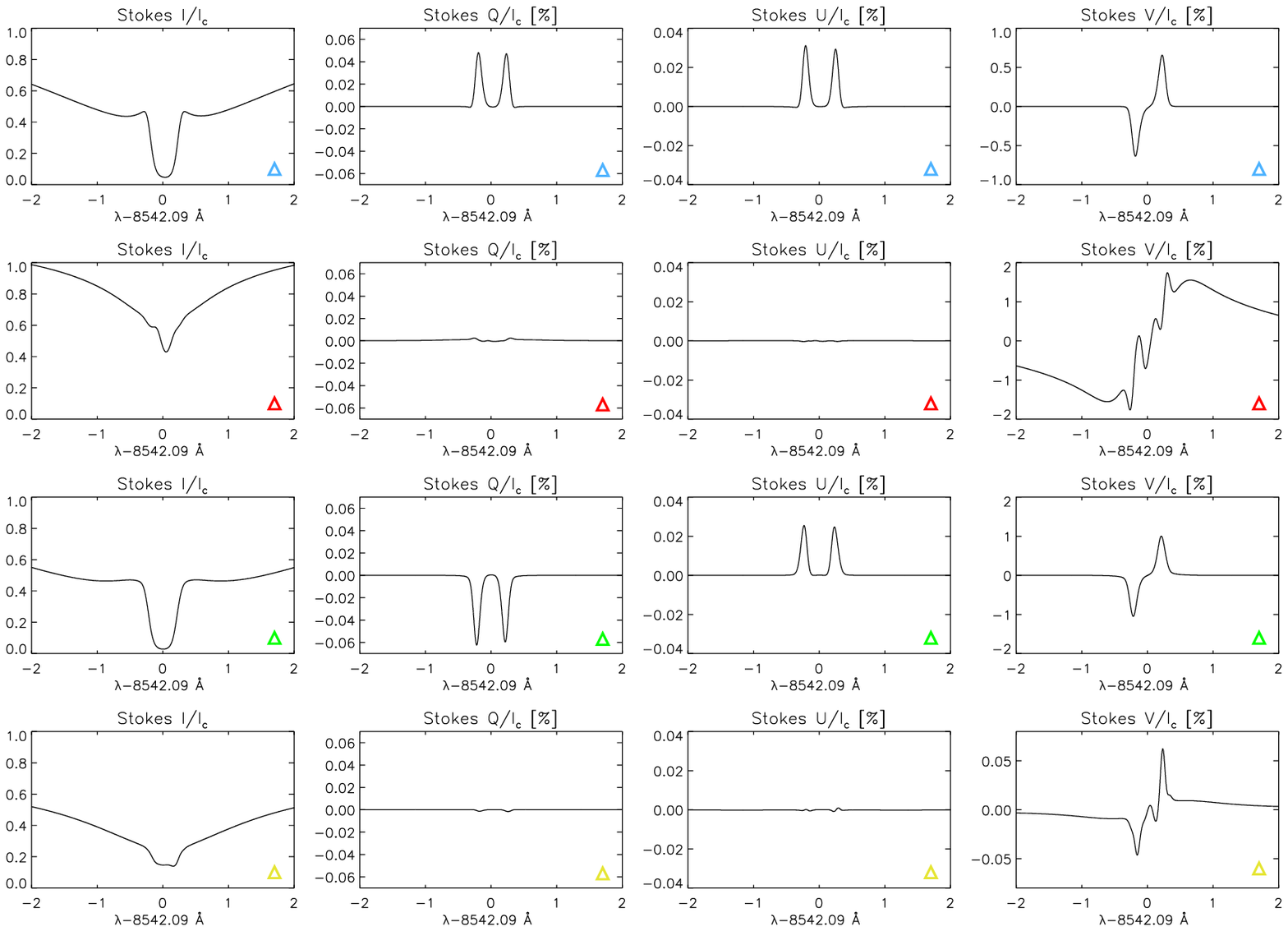}
\caption{Synthetic Stokes profiles from the four atmospheres shown in Figure \ref{uni}. Coloured triangles indicate their location on Figure~\ref{r1}.}
\label{perm3}
\end{figure*}

\section{Synthetic profiles}

We mentioned in section \ref{sims} that the vertical dimension of the simulation extends up to the corona. However, as the Ca~{\sc ii} line is only sensitive to photospheric and chromospheric layers, we chose a reduced height range to perform the synthesis of the Stokes profiles. Thus, we omitted the higher atmospheric layers using a geometrical height that ranges from -0.65 to 2.3~Mm. The selected spectral range covers $\pm$5~\AA \ from the centre of the Ca~{\sc ii} line, i.e. $\lambda=8542.09$~\AA, with a spectral sampling of 1~m\AA. We computed the atomic population densities using {\sc nicole}, assuming statistical equilibrium and under the field-free approximation. Therefore, we do not consider 3D radiation effects in the synthesis of the Stokes profiles and each profile is computed under 1.5D approximation, i.e. each column is treated independently, as if it were infinite in the horizontal direction. However, as stated by \cite{delaCruzRodriguez2012}, this assumption is fairly correct for the case of the Ca~{\sc ii} IR line. Moreover, we included, following the mentioned work, an ad hoc microturbulence to enhance the line core width. We used the same value employed in the previously cited work, i.e. $v_{\rm mic}$=3~km/s. Although there is no justification for adding this microturbulence we need to manually enhance the line width to avoid creating larger, and artificial, polarimetric signals \citep{delaCruzRodriguez2012}. Finally, we synthesised each profile using a macroturbulence value of 1.5~km s$^{-1}$, i.e. we spectrally degraded the Stokes parameters through a convolution with a Gaussian profile with 42 m\AA \ width aiming to simulate real conditions. In this regard, we chose a value slightly larger than that of present solar telescopes. For example, the equivalent width of Hinode/SP spectral point spread function is equivalent to a macroturbulence value of 1.2~km s$^{-1}$ \citep{Lites2013}. We used a single macroturbulence value for degrading the synthetic profiles to reduce the content of the manuscript. If the reader is interested in more details, we strongly suggest to examine the work of \cite{delaCruzRodriguez2012} for a discussion of the effects the instrumental degradation and random noise produce on the polarization signals of synthetic Ca~{\sc ii}.

Finally, we did not include in this work the effect of the isotopic splitting \citep{Leenaarts2014} of the Ca~{\sc ii} 8542 \AA \ line because we believe that is not very important for this study, but we have to bear in mind that it should be included in future analysis, mainly when we perform inversions, to avoid erroneous temperature and velocity gradients.

\subsection{Monochromatic intensity information}

The first study we performed was the comparison of the Stokes $I$ intensity images at different monochromatic wavelengths with the temperature distribution at different heights. We show the results in Figure \ref{sivar} where we can see that the continuum pattern is mostly reproduced, albeit with lower contrast, at $\Delta \lambda=-5$~\AA, see leftmost column. However, we can identify a clear distinction respect to the temperature stratification, i.e. the presence of continuum bright intergranular regions that are co-spatial with strong longitudinal fields. See, for instance, the region at (2,3) Mm in the leftmost bottom panel, where the magnetic field patch has produced a bright region that was completely dark in the temperature map, upper panel. The explanation behind is the so-called \textit{Wilson depression} that produces a shift of the geometrical heights at the same optical depth generating the detection of photons that come from a different atmospheric layer. In our case, as the temperature inside the flux tube monotonically decreases with height, see Figure \ref{uni}, the height shift is towards deeper atmospheric layers. Additionally, we can also find several bright points that correspond to localised longitudinal magnetic field concentrations. Thus, the far wings (almost continuum) of the Ca~{\sc ii} line will appear as intergranular bright points in the presence of moderate/strong magnetic fields.

The bottom panel of the second column of Figure \ref{sivar} shows the intensity values at $\Delta \lambda=-2$~\AA. In this case, the reversed granulation pattern is identifiable and it resembles pretty much what we saw for the temperature spatial distribution, top panel on the same column. The only difference, again, is the presence of bright regions where the longitudinal magnetic field is stronger.

The comparison with the temperature spatial distribution becomes less clear as we examine wavelengths that are closer to the line core. We tentatively select some wavelength locations where the intensity map resembles the different heights of the three rightmost temperature panels. Although it is less clear, we still see a good correlation between the structures of the temperature and the intensity panels, mainly hot structures.

\begin{figure*}
\includegraphics[width=16.0cm]{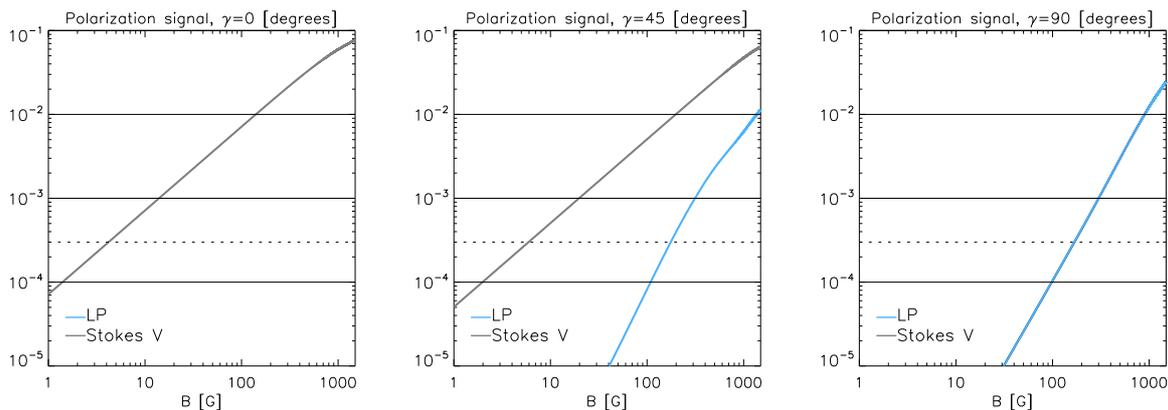}
\vspace{-0.3cm}
\caption{Polarization signals generated by the Stokes profiles synthesised from the FALC model with a magnetic field vector constant with height. Blue colour shows the total linear polarization signals while gray colour depicts the maximum Stokes $V$ signals. Each column corresponds to a different inclination angle and dotted line marks a reference polarization value of $3\times10^{-4}$ of $I_c$.}
\label{max_pol1}
\end{figure*}

\subsection{Selected Stokes profiles}

In this section we aim to study the spectral properties of the Stokes profiles produced by the pixels studied in section \ref{pixelsofinterest}. In this regard, we display in Figure \ref{perm3} the corresponding synthetic Stokes profiles. We identified each pixel including in each panel a coloured triangle following the same colour code used in Figures \ref{r1}, \ref{vertical}, and \ref{uni}. If we focus on the upper row, blue triangle in Figure \ref{r1}, we can see that Stokes $I$ line core is very deep. This is probably due to the lack of temperature at chromospheric layers \citep{Leenaarts2009}. However, the general shape of the intensity profile still resembles the observed atlas profile (see Figure \ref{atlas}). Linear polarization signals are around $3-5\times10^{-4}$ and they show small asymmetries. The reason behind the last property is the jump along the line of sight of the magnetic field stratification, canopy structure. In addition, we can also find large Stokes $V$ signals, indicating that the magnetic field is slightly inclined. Remarkably, the magnetic field vector is almost null below 500 km for this pixel, see Figure \ref{uni}, indicating that these signals are produced by magnetic fields pertaining to the upper photosphere and low chromosphere \cite[see also][]{delaCruzRodriguez2013}. 

Second row shows the Stokes profiles for the pixel located in the centre of the flux tube, red triangle in Figure \ref{r1}. Stokes $I$ shows large intensity values for the core and the wings due to the higher temperature inside the tube. In addition, the line depth is greatly reduced although the line core is still in absorption. Regarding  the linear polarization signals, we found that they are extremely low because the magnetic field vector is vertical. Finally, Stokes $V$ shows a complicated behaviour with signals in the external and broad wings, due to the photospheric magnetic field, and large signals in the line core due to the chromospheric contribution. Moreover, the core wavelengths show several peaks with opposite polarity. We believe that is related to the complex Stokes $V$ response to the longitudinal magnetic field that we show in Figures \ref{rfFig} and \ref{rf2d}.

\begin{figure*}
\centering
\vspace{-0.7cm}
\includegraphics[width=17.5cm]{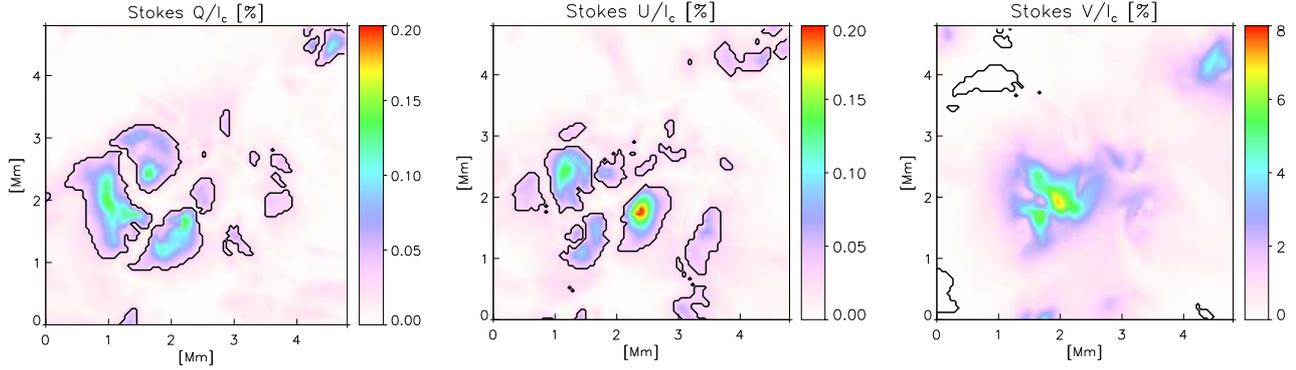}
\vspace{-0.1cm}
\caption{Maximum polarization signals for the region enclosed by the square in Figure \ref{context}. Black contours designate the regions where $Q$ and $U$ are larger than $3\times10^{-4}$ of $I_c$ (left and middle panels) while they mark the regions where the Stokes $V$ amplitude is smaller than $3\times10^{-4}$ of $I_c$ (rightmost panel).}
\label{max_pol2}
\end{figure*}

Third row shows the pixel corresponding to the green triangle in Figure \ref{r1} which behaviour is essentially the same as the pixel of the first row, although it shows opposite polarities for Stokes $Q$. Thus, we found polarization signals generated by a magnetic field that is located at atmospheric layers above 500 km.

Finally, we plot in the last row the synthetic profile from the pixel designated by the yellow triangle in Figure \ref{r1}. In this case, the low magnetic field strength generates very low linear polarization signals, making them very difficult to detect as was already mentioned in \cite{delaCruzRodriguez2012}.

\subsection{Polarimetric signals}

We want to examine in this section the polarimetric signals we can expect from solar observations of regions with moderate magnetic field strength, for example, network regions. In the work of \cite{delaCruzRodriguez2012} the authors focused on the analysis of simulated quiet solar regions emphasising a proper treatment of instrumental degradation and noise. In our case, as we explain in section \ref{sims}, we are going to study the polarimetric signals using an ideal spectral sampling of $\Delta \lambda=$ 1 m\AA \ and we degraded the Stokes profiles with a macroturbulence value of 1.5~km $s^{-1}$, i.e. we convolved each synthetic profile with a Gaussian profile with 42 m\AA \ width.

We first study the maximum polarization signals generated by a simple model, FALC, with constant magnetic field values for all heights. We fixed the magnetic field azimuth to 45 degrees and we changed the magnetic field strength from 0 to 1500~G. We repeated the process using an inclination value of 0, 45, and 90 degrees. The results of this study are shown in Figure \ref{max_pol1} where we plot the maximum linear polarization values (blue), computed as $LP=\sqrt{Q^2+U^2}$,  and the maximum Stokes $V$ signals (gray). We can see that, for measuring the linear polarization produced by transversal fields of 100~G (blue in the middle and rightmost panels), the noise value should be smaller than our reference noise value, i.e. $3\times10^{-4}$ of $I_c$ \citep[see also][]{delaCruzRodriguez2012}. On the contrary, for the same level of noise, we would detect circular polarization signals from longitudinal fields of 10~G (gray colour in the leftmost and middle panels).

However, although the previous results are very useful, we employed more complex atmospheres as the one provided by the {\sc bifrost} code and we computed the maximum polarization signals in the same field of view enclosed by the square of Figure \ref{context}. The corresponding results are plotted in Figure \ref{max_pol2}.  Focusing on Stokes $Q$ and $U$, left and middle panels, we find a considerable amount of linear polarization signal in the borders of the flux tube that produces maximum signals larger than 0.1 per cent ($1\times10^{-3}$) of $I_c$. The reason for these linear polarization signals is the inclined magnetic field lines that fan out with height at the edges of the magnetic structure. Thus, Stokes $Q$ and $U$ signals are produced by inclined upper photospheric and chromospheric fields. In addition, these polarization signals are high enough to be easily detectable for current polarimeters with short integration times. Finally, Stokes $V$ signals are very large in the inner part of the flux tube and they fill almost  all the field of view (note the absence of black contours) with values larger than the reference noise value of $3\times10^{-4}$ of $I_c$.

\section{Summary}

We used the {\sc nicole} code to understand which are the different regions of sensitivity of the Ca~{\sc ii} 8542 \AA \ line to changes of the atmospheric parameters. We have found that the line wings are sensitive to changes of the atmospheric physical parameters that take place between $\log$ $\tau \sim[0,-4]$. Then, we encountered a region, around $\log$ $\tau \sim-4.5$, of low sensitivity to perturbations in all the atmospheric parameters \cite[already mentioned by][]{Pietarila2007}. Above this region, from $\log$ $\tau \sim[-4.5,-5.5]$, response functions are large indicating that small perturbations of the atmospheric parameters will be translated in distinct, albeit complex, spectral features. This means that, measuring the Ca~{\sc ii} 8542 \AA \ line, we will be able to detect variations of the atmospheric parameters from the bottom of the photosphere, $\log$ $\tau=0$, to the middle of the chromosphere $\log$ $\tau \sim-5.5$. 

Later, we used a realistic MHD snapshot to improve our knowledge of the Ca~{\sc ii} 8542 \AA \ line. Although the simulated regions show a large variety of scenarios where we can study the properties of the atmospheric parameters we focused only on a network patch to simplify the analysis. We first examined the physical configuration of the atmosphere to facilitate the understanding of the following sections. After that, we compared the Stokes $I$ profile at different monochromatic wavelengths with the temperature distribution at different heights, finding a good correlation with the spatial distribution of temperatures in the simulation from $Z$=0 to 1~Mm. However, some discrepancies arose in certain spatial locations due to the fact that the simulated temperature sometimes could show very low values at chromospheric layers. 

Then, we focused on the polarimetric profiles. First studying the Stokes profiles produced by selected pixels finding linear polarization signals at the edges of the flux tube produced by canopy structures. In addition, we found a complex Stokes $V$ profile in the centre of the flux tube with multiple lobes in the line core wavelengths that represent the same complex behaviour found in the Stokes $V$ response functions to the longitudinal field. The last pixel examined in detail shows no detectable polarization signals, at least for the noise value we would expect for an observation with short integration times, indicating that the linear polarization signals outside the flux tube structure are scarce \citep[see also][]{delaCruzRodriguez2012}.

In the last study, we analysed the maximum polarimetric signals produced by a simple model (FALC) and by a small magnetic patch of the simulated \textit{enhanced network}. The first case provided us a general idea of how large the magnetic field should be to generate linear and longitudinal polarization signals above a certain noise threshold. Assuming a threshold value of $3\times10^{-4}$ of $I_c$,  we established that linear polarization signals are detectable when the magnetic field strength is larger than 100 G while Stokes $V$ signals are generated above this threshold with magnetic fields larger than 10 G. Regarding the latter study, the network patch generates linear and circular polarization signals at its edges although in the rest of the examined region the linear polarization signals are extremely low, falling below our reference threshold, i.e. $3\times10^{-4}$ of $I_c$.
 
Our results demonstrate that the Ca~{\sc ii} line is able to continuously scan the photosphere and the low/middle chromosphere, being very sensitive to the temperature information and producing low polarization signals when the magnetic field is moderate/strong. We can conclude then, that the Ca~{\sc ii}~8542~\AA \ line presents itself as one of the best candidates for future chromospheric magnetometry.

\section*{Acknowledgements}
The authors acknowledge the helpful comments and ideas B. Ruiz Cobo and H. Socas Navarro provided during the developing of the manuscript. We also want to thank Prof. M. Carlsson for his helpful comments and corrections after reviewing the manuscript. JdlCR is supported by grants from the Swedish Research Council (VR) and the Swedish National Space Board (SNSB).

\bibliographystyle{mnras} 
\bibliography{ca} 

\bsp	
\label{lastpage}
\end{document}